# MorphoCloud: Democratizing Access to High-Performance Computing for Morphological Data Analysis

**A. Murat Maga**[*], Department of Pediatrics, University of Washington, Seattle WA 98195; Center for Developmental Biology and Regenerative Medicine, Seattle Children's Research Institute, Seattle, WA 98101; ORCID: 0000-0002-7921-9018

**Jean-Christophe Fillion-Robin**[*]. Kitware Inc., Clifton Park, NY, 1206; ORCID: 0000-0002-9688-8950

Corresponding author: A. Murat Maga, maga@uw.edu

* Joint first authorship.

## Abstract

The digitization of biological specimens has revolutionized the field of morphology, creating large collections of 3D data, and microCT in particular. This revolution was initially supported by the development of open-source software tools, specifically the development of SlicerMorph extension to the open-source image analytics platform 3D Slicer. Through SlicerMorph and 3D Slicer, biologists, morphologists and scientists in related fields have all the necessary tools to import, visualize and analyze these complex and large datasets in a single platform that is flexible and expandable, without the need of proprietary software that hinders scientific collaboration and sharing.

Yet, a significant "compute gap" remains: While data and software are now open and accessible, the necessary high-end computing resources to run them are often not equally accessible in all institutions, and particularly lacking at Primarily Undergraduate Institutions (PUIs) and other educational settings. Here, we present MorphoCloud, an "IssuesOps"-based platform that leverages Github Actions and the JetStream2 cloud farm to provide on-demand, research-grade computing environments to researchers working with 3D morphological datasets. By delivering a GPU-accelerated full desktop experience via a web browser, MorphoCloud eliminates hardware barriers, enabling complex 3D analysis and AI-assisted segmentation. This paper explains the platform and its architecture, as well as use cases it is designed to support.

# Introduction

## The Harnessing 3D Data in Biological Sciences

The last decade has witnessed a paradigm shift in biological data collection. Over the years, the NSF Biology Directorate has made significant investments in the 3D digitization of biological specimens held in natural history collections. Initiatives like oVert [1] (OpenVertebrates) and MorphoSource [2] have generated and published hundreds of terabytes of high-resolution 3D data, ranging from surface models to high resolution Computed Tomography (CT) scans. These projects have successfully democratized *access to data*, ensuring that a researcher in a remote location or a rural institution can download the same digital specimens as a curator at a major natural history museum.

## Bridging the Software Gap: SlicerMorph

While the digitization of specimens removed physical barriers, a significant "software barrier" remained. Historically, interacting with 3D biological data required expensive, proprietary visualization software. Furthermore, the workflows were fragmentary, necessitating serial use of 4-5 different software packages to accomplish the tasks of importing data from scanners, visualizing, segmenting and analyzing them. This fragmented workflow with incompatible formats and proprietary software created significant data management challenges, as the same specimen was being represented in 4-5 different formats and data representation in various stages of the pipeline. These formats were incompatible with each other, making going back and forth through the workflow to change and improve very difficult, if not impossible.

To overcome these challenges we devised the SlicerMorph project [3]. 3D Slicer (aka Slicer) was chosen as the primary platform to develop the extension because Slicer was already a very capable application, providing in a single platform visualization, processing and analysis of 3D volumetric data [4], [5]. Missing functionality could be added on thanks to its open-source and modular nature. SlicerMorph extended the existing core functionality of Slicer with generic (ImageStack) and vendor specific (SkyscanReconImport, GEVolImport) functionality to import imagestacks from microCT and other modalities; implemented the Generalized Procrustes Analysis (GPA) for visualizing shape variation and other geometric morphometrics modules (e.g., PseudoLMGenerator, PlaceGridLandmarks, ProjectSemiLMs), as well as fully automated landmarking tools like ALPACA and MALPACA [6], [7], [8], [9]. Within in the SlicerMorph extension, there are additional modules that provide convenience functions such as segmenting endocranial space in mammalian skulls (SegmentEndocranium), alignment and registration of 3D models (FastModelAlign), acquiring high-resolution 3D renderings (HiResScreenCapture). SlicerMorph team continues to work with the other Slicer developers to incorporate changes into the core functionality of the Slicer, such as the new markups infrastructure that enabled landmarking functionality, help diagnose and resolve issues regarding using and interacting large datasets in Slicer including rendering, segmentation and saving, as well as adding new functionalities to (e.g., ColorizeVolume). At this juncture, all fundamental digital morphology tasks such as importing, visualizing, segmenting, and analyzing can be done completely within Slicer. For domain specific inferential statistics on morphometric data, SlicerMorphR R library provides convenient import of SlicerMorph's GPA results into the R ecosystem with a single line of code.

At the same time SlicerMorph project has grown from a single extension to an ecosystem of integrated extensions that work together for complex workflows. For example, now SlicerMorph's Photogrammetry extension [10], [11] allows converting large number of photographs into 3D textured models, which can then be landmarked using the existing tools in Slicer, which in return can be fed into SlicerMorph's Dense Correspondence Analysis (DeCA) toolkit that simultaneously create 3D template of models and sample dense semi-landmarks [12], [13], and finally the new output can be can be analyzed with the GPA module in the SlicerMorph extension proper. Mouse Multi Organ Segmentation (MEMOs) extension uses deep-learning to rapidly segment 50 structures diceCT scans of mouse embryos [14]. Segments generated by MEMOs can be immediately fine-tuned with the Slicer's Segment Editor. ScriptEditor extension allows taking example python codes from Slicer's Script Repository and tweaking it for the user's specific needs all inside the Slicer's environment with support for code auto complete and function discovery and help documentation. The emerging MorphoDepot extension allows teams, whether they are geographically separate collaborators, or instructor and students in a class, to work jointly on the same segmentation using the well-established fork-and-contribute model of open-source development.

All in all, it is now possible to conduct all digital morphology tasks entirely within the Slicer platform, and maintain full control over the workflow. Any unique functionality that is currently not available can be implemented rapidly, thanks to the growing number of AI-assisted coding platforms that are trained on the fully open Slicer source code, documentation and its Application Programming Interface (API).

## The Compute Gap

Despite the availability of open data and open software, a final barrier prevents the full democratization of this science: the *hardware* gap. While the application can be free, the computational cost of running it is not. Processing high-resolution microCT scans requires workstations with high-end GPUs and large amounts of RAM. Particularly for segmentation a typical rule of thumb is 4-8 times more physical memory than the size of the dataset. A "typical" high-resolution microCT scan can easily be 2048 voxels (or more) in each dimension. If the intensities in this image are represented by 16-bit data, it means simply importing this data as a 3D array would require 16GB of physical memory. During the segmentation, necessities of having multiple copies of this array mean transient memory usage can be as large as 60-100GB, which often exceeds resources on personal computing devices. This large resource requirement, coupled with the need to have data-center grade GPUs to benefit from emerging AI technologies to process and segment large microCT datasets creates a "digital divide." Research-intensive universities often have high performance computing environments, which can be leveraged to accomplish these tasks by their faculty and students. Also, often there are shared computer labs with relatively modern and powerful workstations, since computationally heavy courses are fairly common in such institutions. On the other hand, Primarily Undergraduate Institutions (PUIs) and other teaching focused institutions may lack such resources. An instructor who wants to incorporate existing 3D morphological data to supplement their curriculum, not only tasked with identifying which datasets to use, but where and how students will access these resources.

## The MorphoCloud Mission

ACCESS is a program established and funded by the U.S. National Science Foundation to help researchers and educators, with or without supporting grants, to utilize the nation's advanced computing systems and services, at no cost [15]. However, most of the resources within the ACCESS program are traditional high-performance computing environments that are set to be used as queue-based batch work systems and are not conducive to the interactive computing that is necessary for digital morphology. Alternative to these HPC systems are cloud farms that provide virtual machines that come in many different configurations and can be "rented" on demand.

While cloud farms may offer cost-effective solutions to "rent" powerful computers without investing significant sums of capital equipment, they have their own challenges. These virtual computers need to be configured from the bare operating system, meaning the "tenant" needs to configure the network, install all the software they need and maintain it securely. The second challenge is, commercial cloud farms charge rent by the minute the virtual computer is online, regardless of if the computer is actively being used or simply sitting idle. So, any computer that is not being used needs to be turned off and "shelved" to avoid charges. Any "run-away" or "stray" virtual machine has the potential to rack up charges in thousands of dollars (or exhaust the educational credits) until it is discovered and turned off. While there are orchestration tools and technologies to avoid such situations, these are jobs for full-time system administrators, which are hard to come by at most PIUs and other non-research focused academic institutions, and are not tasks most morphologists or researchers in related fields are familiar with or have the bandwidth to acquire the skills.

JetStream2 is a public cloud farm that is part of the ACCESS program funded by NSF [16]. Like any resource in the ACCESS program, any US researcher can request allocation on the JetStream2 at no cost. However, they still need to accept the responsibility to provision their own instances with the software, and keep tabs on their running instance not to exhaust their resources. So, the challenges associated with commercial cloud vendors still apply to JetStream, only the cost aspect of it is changed.

Here we introduce the MorphoCloud, an "IssuesOp" based platform built on the JetStream2 and allows users to provision and manage powerful remote workstations (aka instances) with simple commands [17]. Our aim is to bridge the compute gap by providing "one-click" access to modern, research-grade computing infrastructure without requiring users to become system administrators. Created instances can be accessible via web browser and already come with the Slicer and SlicerMorph ecosystem preinstalled, enabling research and teaching activities to commence at the very moment the instance is online.

## System Design & Architecture

### Philosophy: IssuesOps

The core design philosophy of MorphoCloud is "IssuesOps"—a command-driven interface. MorphoCloud uses the Github Issues as the primary means to request, get approved and control the instances. Users interact with the system by typing natural-looking commands (e.g., `/create`, `/unshelve`, `/renew`) as comments on an issue. This approach provides a simple, collaborative, and persistent record of all interactions. There are only two requirements to use MorphoCloud instances: A public ORCID, and a Github account, both of which are free to acquire.

### Overview of MorphoCloud Technology Stack

MorphoCloud uses existing, established open-source projects to facilitate the functionality. These are:

- **Frontend** (User Interface): GitHub Issues. The issue tracker serves as the dashboard, where users request resources and receive status updates via bot comments.
- **Orchestration**: GitHub Actions. These workflows parse user comments, validate permissions, and execute the necessary logic. Crucially, the workflows run on a Self-Hosted Runner that resides within the JetStream2 cloud network, allowing direct, secure communication with the OpenStack API.

- **Backend** (Compute): OpenStack running on JetStream2. This provides the raw virtualization layer.
- **Provisioning**: Cloud-Init and Ansible. When an instance is spawned, a custom user-data script bootstraps the environment. It installs Ansible in a virtual environment and pulls a specific configuration playbook from the Exosphere repository customized for the MorphoCloud to set up the remote desktop environment.
- **Delivery**: Apache Guacamole. A clientless remote desktop gateway that renders the desktop environment to HTML5. This allows users to access their high-performance instance via any modern web browser, bypassing local hardware limitations.

## Security Model

Security is managed through a "vendorized" workflow approach. The complex logic resides in a central repository (MorphoCloudWorkflow), while user projects (eg. MorphoCloudInstances) import these workflows. Sensitive credentials (SSH keys, OpenStack API tokens) are stored as GitHub Repository Secrets, never exposed in the issue comments, and only accessible to repository owners. Access control to the managed instances is enforced via a defined list of administrators who must approve instance creation and management requests using the `/approve` command.

This flexible design allows anyone who obtains their own allocation on the JetStream2 to fork the MorphoCloudWorkflow repository, customize it for their own use cases (e.g., different software tools and/or user interface) and deploy it for their own community. Step-by-step instructions for creating custom deployments similar to MorphoCloud can be found on the MorphoCloudWorkflow repository [18].

## Day to Day Operations

This is the lifecycle of a MorphoCloud instance from an individual user's perspective:

1. **Request and Approval:** User opens a new Issue using a standardized template identifying themselves via their ORCID profile, briefly explaining their project. The MorphoCloud team reviews the request to make sure the request fits in the MorphoCloud intended usage and approves it often within a few hours. If there are concerns or clarifications needed, more comments are added to the user's specific github issue.
2. **Instance Initialization**: An approved user can initialize the process by issuing `/create`, command, which starts provisioning the instance with all the necessary software, along with the user's persistent storage volume. The issue page displays each provisioning stage so that the user can see the progress (or notice errors).
3. **Notification**: Once an instance is ready and online, the system indicates that provisioning is completed, and sends the user an email with access credentials.
4. **Access to remote desktop**: User can access the instance via a web browser by clicking the access URL and entering the provided credentials. Alternatively, the user can connect to the instance via the VNC protocol for a slightly more performant and more native-like experience. Command line access to the instance via SSH protocol is also a possibility.
5. **Shelving and Unshelving:** An active instance stays online up to four hours, unless the user chooses to further extend their session. Users can extend their sessions four hours at a time by clicking an icon on the remote desktop. An inactive remote desktop session will "shelve" itself automatically at the end of the last four-hour extension. Alternatively, when done, the user can shelve their own instance by typing the command `/shelve` on their issue's page. A shelved instance can be made online again by entering the `/unshelve` command on the issue page. Unshelving is much faster than fully provisioning the instance and often the instance is online within a few minutes after the command is entered. A new email with access credentials is sent to the user each time as those may change after shelving and unshelving. If the credential email is not received or lost for any reason, it can be re-requested via `/email` command.

6. **Instance expiration and renewal:** Each instance has a life span of 60 days, after which the instance expires and gets deleted automatically. 60-day countdown starts from the time /create command is issued by the user. As the life span is nearing, a remainder email is sent to the user about the upcoming expiration. The user then can issue the `/renew` command to extend the life span for another 60 days. If the user takes no action, both the instance and their persistent storage volume will be deleted at the end of the 60 days.

7. **Additional commands for troubleshooting:** Due to reasons beyond our control (e.g., networking lost during provisioning, faulty host or GPU), instances may be left in an inaccessible state. In such cases, additional commands such as `/delete_instance` can be issued by the user to purge the instance and reprovision their own instance by reissuing the `/create` command. This is often the simplest and fastest solution to regaining access and is harmless since this command only removes the instance and the user's persistent storage volume is untouched.

To support instructors offering short courses and workshops centered on the SlicerMorph ecosystem, MorphoCloud also includes a "Workshop" instance request type. The primary purpose of the workshop instance type is to bulk provision many instances and distribute the access credentials to participants prior to the event. Moreover, these instances do not automatically shelve themselves after four hours of inactivity, but remain online continuously. However, their life span is much shorter. Workshop instances can only be requested for up to five days (plus 24h of grace period for setup). The course organizer makes the request on behalf of participants, receives the access credential emails for all instances created for the event, and is responsible for distributing them to the participants.

There are few things to emphasize. After the approval, instance creation is not automatic; as explained above, the user needs to initiate that via the /create command. Provisioning an instance from is similar to configuring a brand-new computer from scratch; it takes time. The time from issuing the `/create` command and the instance becoming online is at least 15 minutes and can be substantially longer. The primary factor that affects the wait time is the number of tasks in the queue that needs to happen before the system can get to the specific /create command. Given that it currently takes about 15 minutes to provision an instance, if 4-5 users issue the /create command more or less at the same time, it may take more than an hour for the last instance to be created. This is why there is a 24h grace period to the workshop. Our suggestion is the event organizer provision the instances the day before the event to have ample time for all the provisioning to take place.

## Data management

As briefly mentioned above, the instances and where the user data is stored is decoupled by design. Instances are considered "ephemeral" and as such can be deleted and recreated to troubleshoot or to upgrade the base software tools deployed. Therefore we advise the MorphoCloud users not to store any critical data on the boot disk of the instance, but instead store everything in their private, persistent storage volume. This 100GB volume is mapped as an external disk under the `/media/volume/MyData`. Primary document root for Slicer and other applications also points to this folder.

Apache Guacamole interface (aka the web browser connection) provides a convenient utility to traverse the remote file system and to download individual files by double clicking them. Similarly, while using the Guacamole interface, if the user drag and drop a file from their own local computer to the remote desktop, the file will be automatically uploaded to the remote computer and will be placed under the persistent storage. For bulk data ingress and egress onto the instance using standard SFTP and SCP protocols is suggested.

## Discussion

MorphoCloud is designed to be used by people who need intermittent access to powerful computers to process 3D datasets interactively and cannot justify investing significant capital expense for sporadic usage of a technology that gets depreciated and outdated quickly. It should be noted that because it uses the shared, national resource (JetStream2) as the backend, there is no guarantee that at a specific time of the day, there will be sufficient resources available for everyone. In formal teaching settings like academic classes, MorphoCloud is best utilized as an asynchronous tool to supplement the curricular activities, as there is no reservation system on the JetStream2 to allow pre-reserving instances for a specific time for everyone in that class.

Where MorphoCloud is particularly useful is intense short courses and workshops focused on digital morphology and related fields, as it isolates a lot of the technological complexity associated with running such events. Traditionally in such events participants are expected to bring their own computing devices. Undoubtedly these computers are a heterogeneous mixture of operating systems, hardware configurations and restrictions (e.g., devices centrally managed by IT departments). The time spent on troubleshooting one-off challenges faced by the participants can easily derail the schedule. Because MorphoCloud provides an identical environment to all users, once instructors run their workflow on their instances, they can be confident that it will work for everyone as intended. For such events, instructors can provision all the nodes the day before the activity, distribute the access credentials to the user on the day of the event, and because the only software requirement from the user's side is a web browser, the event can proceed in a timely fashion. Because workshop instances do not auto-shelve themselves in time of activity, once they are provisioned, they are guaranteed to stay online throughout the event. However, again there is no procedure guarantee sufficient resources will be available at the time of provisioning. That's why the workshops are given a 24h grace period to launch their instances. In our experience the default g3.l flavor of MorphoCloud is often consistently available for events up to 35 participants. Another issue that such event organizers must be mindful of is that JetStream2 engineering team have scheduled and (occasionally) emergency maintenance of the system. Depending on the type of the maintenance the system might be partially or completely unavailable to the users. We advise checking the JS2's planned outage calendar, or getting in touch with the admins, before planning such an event.

Another use case MorphoCloud shines is using AI-assisted technologies in digital morphology. While pre-trained deep-learning networks—ranging from fully automated pipelines to interactive, promptable models—offer transformative potential for processing CT data, their practical implementation is hindered by a "multi-layered cake" of software dependencies. The deployment of open-source AI tools requires a precise alignment of several distinct architectural layers. At the foundation, the hardware-driver interface (e.g., NVIDIA CUDA) must be exactly compatible with the computational frameworks (such as PyTorch or TensorFlow). Above this, a complex lattice of mathematical and image-processing libraries must be version-matched to the specific segmentation tool. As such, just obtaining the right environment to test a new AI tool can itself be a large time sink. MorphoCloud mitigates these complexities by providing the correct and current software stack that is tested with the Slicer. By hosting these "layered cakes" in a pre-configured, cloud-based infrastructure, often all the user needs to do is to install the extension from the Slicer's catalogue, and test its functionality in a few minutes.

MorphoCloud demonstrates that accessibility is about more than just open data; it requires open infrastructure. By abstracting away the complexity of cloud computing, we empower researchers at under-resourced institutions to participate fully in the era of big data biology.

## Acknowledgments

MorphoCloud and its related services are created by funding from the U.S. National Science Foundation (DBI/2301405) and the National Institutes of Health (HD104435). The platform runs on cyberinfrastructure made available by the U.S. National Science Foundation: Jetstream2 (OAC/2005506) and Exosphere (TI/2229642) through allocation BIO180006 from the Advanced Cyberinfrastructure Coordination Ecosystem: Services & Support (ACCESS) program, also supported by U.S. National Science Foundation (2138259, 2138286, 2138307, 2137603, and 2138296). Initial development of the SlicerMorph extension was supported by the U.S. National Science Foundation (DBI/1759883).

# Tables

**Table 1**. Available instance types, specifications and use cases.

| Type | RAM (GB) | CPU Cores | GPU (A100) | MyData | USE CASES |
|---|---|---|---|---|---|
| g3.l | 60 | 16 | 20GB | 100GB | General purpose morphology and morphometrics |
| g3.l | 125 | 32 | 40GB | 100GB | Photogrammetry, NNInteractive, other AI applications |
| m3.xl | 250 | 64 | --- | 100GB | Computationally intense tasks that don't require GPU |
| r3.l | 500 | 64 | --- | 100GB | Applications requiring large amount of RAM |
| r3. xl | 1000 | 128 | --- | 100GB | Applications requiring large amount of RAM and cores |

**Table 2**. List of most commonly used commands as "IssuesOps"

| COMMANDS | FUNCTION |
|---|---|
| /create | Provision an instance from scratch (one time only, unless its deleted) |
| /shelve | Suspend a running instance (after 4h, instances auto-suspend) |
| /unshelve | Bring a shelved instance online again |
| /renew | Extend the instance by another 60 days (one time only) |
| /email | Resend access credentials (only valid if instance is online) |
| /delete_instance | Delete an instance (used only as a troubleshooting step) |

**Table 3.** List of applications installed on the MorphoCloud. Additionally, users have admin rights on their own instances.

- **3D Slicer** (with the possibility of installing more its extensions)
- **SlicerMorph Ecosystem and other extensions:**
    - **SlicerMorph**: (ImageStacks, GPA, ALPACA, PseudoLMGenerator, FastModelAlign and others)
    - **DeCA**: Morphometrics via dense correspondence analysis.
    - **Photogrammetry**: Generate textured 3D models from photographs
    - **MorphoDepot**: Collaborative segmentation workflows
    - **MEMOs**: AI based organ segmentation for E15 mouse embryos
    - **ANTsPy**: ANTs diffeomorphic image registration and analysis library
    - **NNInteractive**: AI assisted promptable interactive segmentation
    - **PyTorch**: GPU accelerated tensor library for AI tools.
- **R/Rstudio**
- **Python3**

# Figures

**Figure 1.** Apache Guacamole (Web browser access) interface. **A:** Side panel allows copy/paste text into the remote session, browse and download files on the remote server and adjust screen zoom levels. **B:** Shortcuts to commonly used applications and to **MyData** storage volume.

**C:** Displays list of available applications (searchable) **D:** Right mouse click brings desktop context menu, including changing screen resolution and other visual desktop settings.

**E:** Extend the current session for more 4 hours (can be used repeatedly).

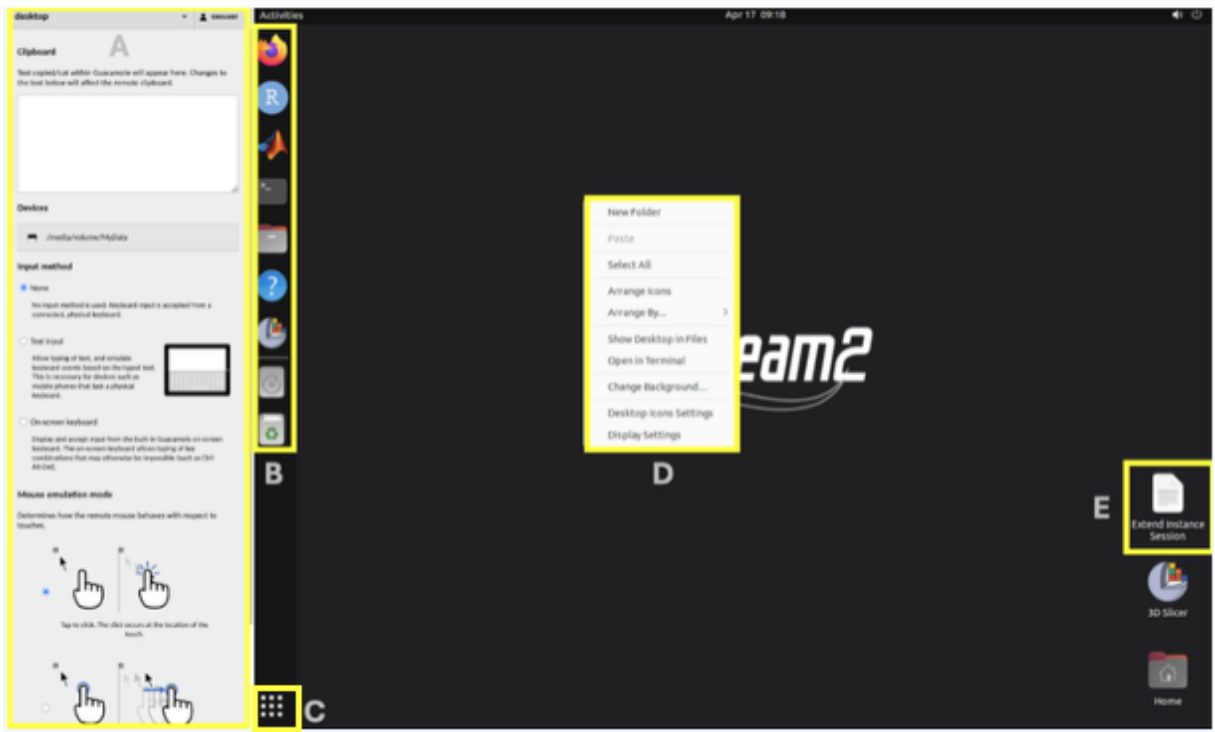

**Figure 2.** Screenshots of MorphoCloud instances running different Slicer extensions. **A.** Rendering of textured 3D model reconstructed through SlicerMorph Photogrammetry extension, with landmarks created via PlaceLandmarkGrid module in SlicerMorph extension. **B.** 3D Volume rendering of a stained bumblebee on a g3.xl instance. Dataset is 1824x1836x2023 voxels (16-bit) and is 12.5GB in size. **C.** Interactive segmentation of molar tooth row as separate segments using NNInteractive extension. User prompts the tool for the regions to keep and regions to remove in a segment. The result shown required 8 prompts (clicks) and took less than 2 minutes to generate. **D.** MorphoDepot collaborative segmentation framework (search results).

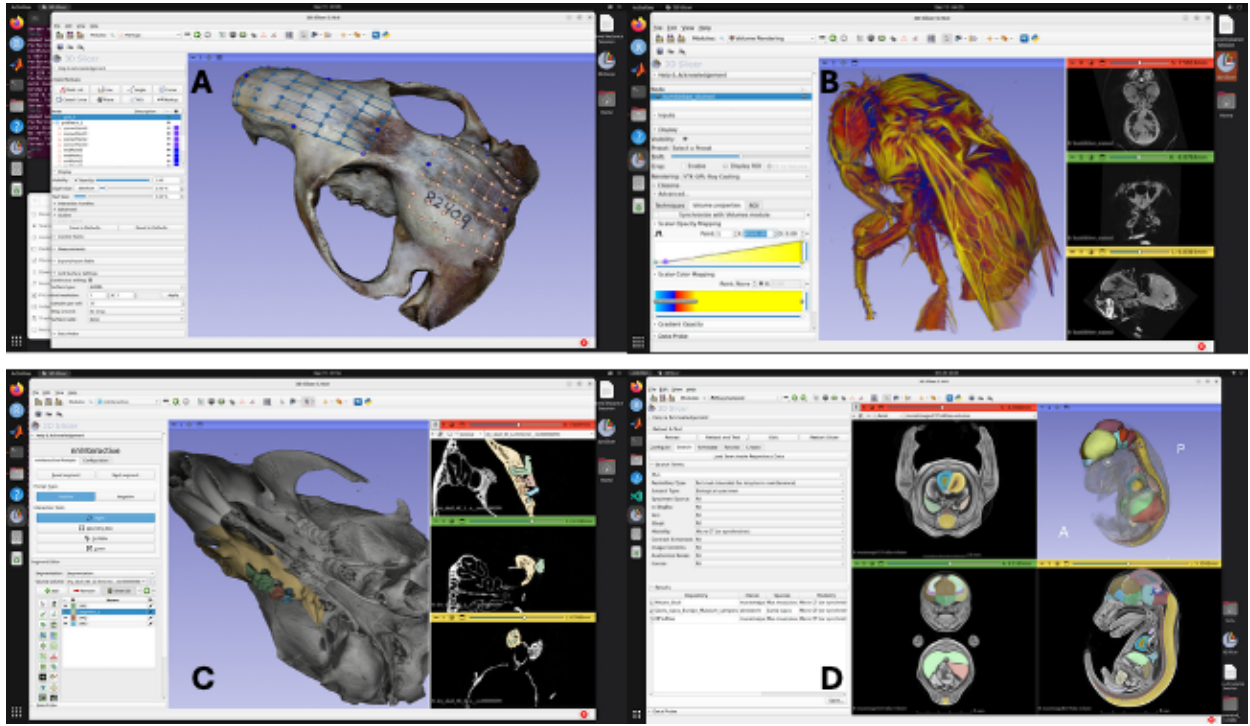